\documentclass[APA,STIX1COL]{WileyNJD-v2}

\pdfoutput=1

\articletype{}%

\received{26 April 2016}
\revised{6 June 2016}
\accepted{6 June 2016}

\usepackage{bm}
\usepackage{textgreek}
\usepackage{units}
\usepackage{xspace}
\usepackage{afterpage}

\usepackage[
  capitalize,           
]{cleveref}             

\usepackage{tikz}
\usepackage{amsmath}
\usepackage{mathtools}
\usepackage{dirtytalk}
\usepackage{derivative}
\usepackage{threeparttable}
\usepackage{lscape}
\usepackage{float}
\usepackage{booktabs}
\usepackage{threeparttable}
\usepackage{caption}
\usepackage{pgfplots}
\usepackage{subcaption}
\usepackage[short,nocomma]{optidef}

\usetikzlibrary{plotmarks,shapes}
\usetikzlibrary{arrows.meta, arrows}
\usepgfplotslibrary{patchplots}
\pgfplotsset{try min ticks=5}
\pgfplotsset{scaled y ticks=false}
\pgfplotsset{compat=newest}

\newlength\figHsmall
\newlength\figWsmall
\newlength\figHbig
\newlength\figWbig
\graphicspath{{figures/}}

\newcommand*{\MAT}[1]{\ensuremath{\bm{#1}}}

\newcommand*{\diff}{\mathop{}\!\mathrm{d}}	
\newcommand{\tran}{{\mkern-1.5mu\mathsf{T}}} 

\renewcommand*{\P}{\MAT P\xspace}

\renewcommand*{\S}{\MAT S\xspace}

\makeatletter
\NAT@longnamesfalse
\makeatother

\begin{document}

\title{Towards a modeling, optimization and predictive control framework for fed-batch metabolic cybergenetics}

\author[1]{Sebastián Espinel-Ríos}

\author[2]{Bruno Morabito}

\author[3]{Johannes Pohlodek}

\author[1]{Katja Bettenbrock}

\author[1]{Steffen Klamt}

\author[3]{Rolf Findeisen}

\authormark{Espinel-Ríos, Morabito, Pohlodek, Bettenbrock, Klamt, Findeisen}

\address[1]{\orgdiv{Analysis and Redesign of Biological Networks, Max Planck Institute for Dynamics of Complex Technical Systems}, \orgaddress{\city{Magdeburg},
\country{Germany}}}

\address[2]{\orgname{Yokogawa Insilico Biotechnology GmbH}, \orgaddress{\city{Stuttgart},
\country{Germany}}}

\address[3]{\orgdiv{Control and Cyber-Physical Systems Laboratory},
\orgname{Technical University of Darmstadt}, \orgaddress{\city{Darmstadt},
\country{Germany}}}

\corres{\email{rolf.findeisen@iat.tu-darmstadt.de}}


\abstract[Abstract]{Biotechnology offers many opportunities for the sustainable manufacturing of valuable products. The toolbox to optimize bioprocesses includes \textit{extracellular} process elements such as the bioreactor design and mode of operation, medium formulation, culture conditions, feeding rates, etc. However, these elements are frequently insufficient for achieving optimal process performance or precise product composition. One can use metabolic and genetic engineering methods for optimization at the intracellular level. Nevertheless, those are often of \textit{static} nature, failing when applied to \textit{dynamic} processes or if disturbances occur.
Furthermore, many bioprocesses are optimized empirically and implemented with little-to-no feedback control to counteract disturbances. The concept of cybergenetics has opened new possibilities to optimize bioprocesses by enabling online modulation of the gene expression of metabolism-relevant proteins via external inputs (e.g., light intensity in optogenetics). Here, we fuse cybergenetics with model-based optimization and predictive control for optimizing dynamic bioprocesses. To do so, we propose to use dynamic constraint-based models that integrate the dynamics of metabolic reactions, resource allocation, and inducible gene expression. We formulate a model-based optimal control problem to find the optimal process inputs. Furthermore, we propose using model predictive control to address uncertainties via online feedback. We focus on fed-batch processes, where the substrate feeding rate is an additional optimization variable. As a simulation example, we show the optogenetic control of the ATPase enzyme complex for dynamic modulation of enforced ATP wasting to adjust product yield and productivity.}



\maketitle


\newpage
\clearpage
\section{Introduction}
\label{sec:introduction_motivation}
The demand for sustainable biotechnological products has grown significantly in recent years \citep{vertes_global_2020,wohlgemuth_bioeconomy_2021}. Although several bioprocesses are commercially successful \citep{sanford_scaling_2016, jullesson_impact_2015}, many are still discarded at early stages because they are not as competitive as traditional technologies. A natural question that arises is how bioprocesses'  efficiencycan be optimized.

The toolbox of bioprocess optimization at the \textit{extracellular} or macro-level includes selection of the bioreactor mode of operation, bioreactor design, optimization of cultivation conditions (pH, temperature, etc.), formulation of culture media, determination of optimal feeding profiles and initial concentrations, among others (cf. e.g. \cite{vandermies_bioreactor-scale_2019,behera_bioprocess_2019,azimi_optimization_2019}). These optimization strategies can influence the overall cell metabolism. Still  they \textit{alone} tend to fail at targeting specific metabolic elements, such as key metabolic fluxes, without affecting other cell functionalities.

Dynamic model-based optimization and predictive control strategies can be used to exploit the dynamic potential of bioprocesses. Dynamic optimization allows finding the optimal dynamic operation conditions, e.g., in \cite{jabarivelisdeh_improving_2016,jabarivelisdeh_model_2018,ryu_model-based_2019,nimmegeers_interactive_2018,del_riochanona_comparison_2019}). Feedback control schemes, especially predictive control approaches as in \cite{jabarivelisdeh_optimization_2018,morabito_multi-mode_2019,jabarivelisdeh_adaptive_2020,morabito_towards_2021,morabito_efficient_2022}, allow one to counteract unknown disturbances such as changes in feed conditions or non-modeled dynamics, while maximizing the production efficiency and rendering a consistent process performance.

At the \textit{intracellular} or micro-level, the bioprocess optimization toolbox includes metabolic and genetic engineering methods for rewiring metabolic pathways. Classical \textit{static} metabolic engineering aims at increasing the cell's product yield, often at the expense of lower biomass yield as the substrate flux diverges from biomass-producing reactions to the product-of-interest pathway. This inevitably decreases the volumetric productivity rates in batch-type bioreactors \citep{venayak_engineering_2015, lalwani_current_2018}. Furthermore, designing dynamic processes based on static metabolic control principles, usually derived under steady-state assumptions, can lead to metabolic imbalances \citep{cui_multilayer_2021}.

Inducible expression of metabolism-relevant proteins via external inputs has emerged as a promising \textit{dynamic} degree of freedom for bioprocess optimization at the micro-level \citep{shen_dynamic_2019,lalwani_current_2018,hartline_dynamic_2021}. Of increasing popularity is the application of optogenetics, the use of light to modulate gene expression \citep{hoffman_optogenetics_2022}. With optogenetics, one can switch on/off fluxes along metabolic pathways via modulation of enzyme expression \citep{tandar_optogenetic_2019,lalwani_optogenetic_2021,zhao_optogenetic_2021}. One can also directly influence cell growth via modulation of the expression of (anti)toxin proteins \citep{lalwani_populations_2021}. In the latter optogenetic examples, the \textit{optimal} light input values were determined using factorial experiments and similar heuristic approaches, resulting in, e.g., two-stage or three-stage fermentations. Furthermore, the inputs were applied in an open-loop fashion, i.e., without online feedback or corrective actions. Considering the often present stochasticity of gene expression \citep{de_vrieze_stochasticity_2020} and the possible presence of process disturbances and batch-to-batch variability, bioprocesses operated in this manner may portray poor reproducibility, moderate-to-poor product quality and a higher risk of failure.

Motivated by these challenges, some authors have proposed \textit{cybergenetic} schemes whereby computer-aided feedback control is used to compensate for uncertainties. In such cases, the corrective actions are calculated outside the cell, e.g., by a computer-aided controller \citep{hsiao_control_2018,khammash_cybergenetics_2022}. To the best of our knowledge, the \textit{biotechnological} applications of cybergenetics have been so far limited to, for instance, controlling the expression of fluorescence proteins and growth-regulatory proteins (e.g., enzymes involved in essential amino acid synthesis or antibiotic-resistance conferring proteins) via optogenetics \citep{milias-argeitis_automated_2016,gutierrez_mena_dynamic_2022}. 

It is believed that the next step in this direction, bearing considerable potential, is to implement \textit{metabolic} cybergenetic systems, i.e., ephasizing dynamic metabolic engineering applications \citep{carrasco-lopez_optogenetics_2020}. Therefore, we seek to extend the scope of cybergenetics to scenarios where metabolic fluxes are to be dynamically manipulated (e.g., towards maximizing the volumetric productivity, achieving a target product yield, rendering a given ratio of products, etc.), while being able to compensate for disturbances and process changes. We aim to use model-based optimization and predictive control methods to exploit the full potential of metabolic cybergenetic systems, considering both cybergenetic inputs and traditional process inputs such as feeding rates simultaneously.

\begin{figure*} [htb]
\begin{center}
\includegraphics[trim={0 3cm 0 0},scale=0.4]{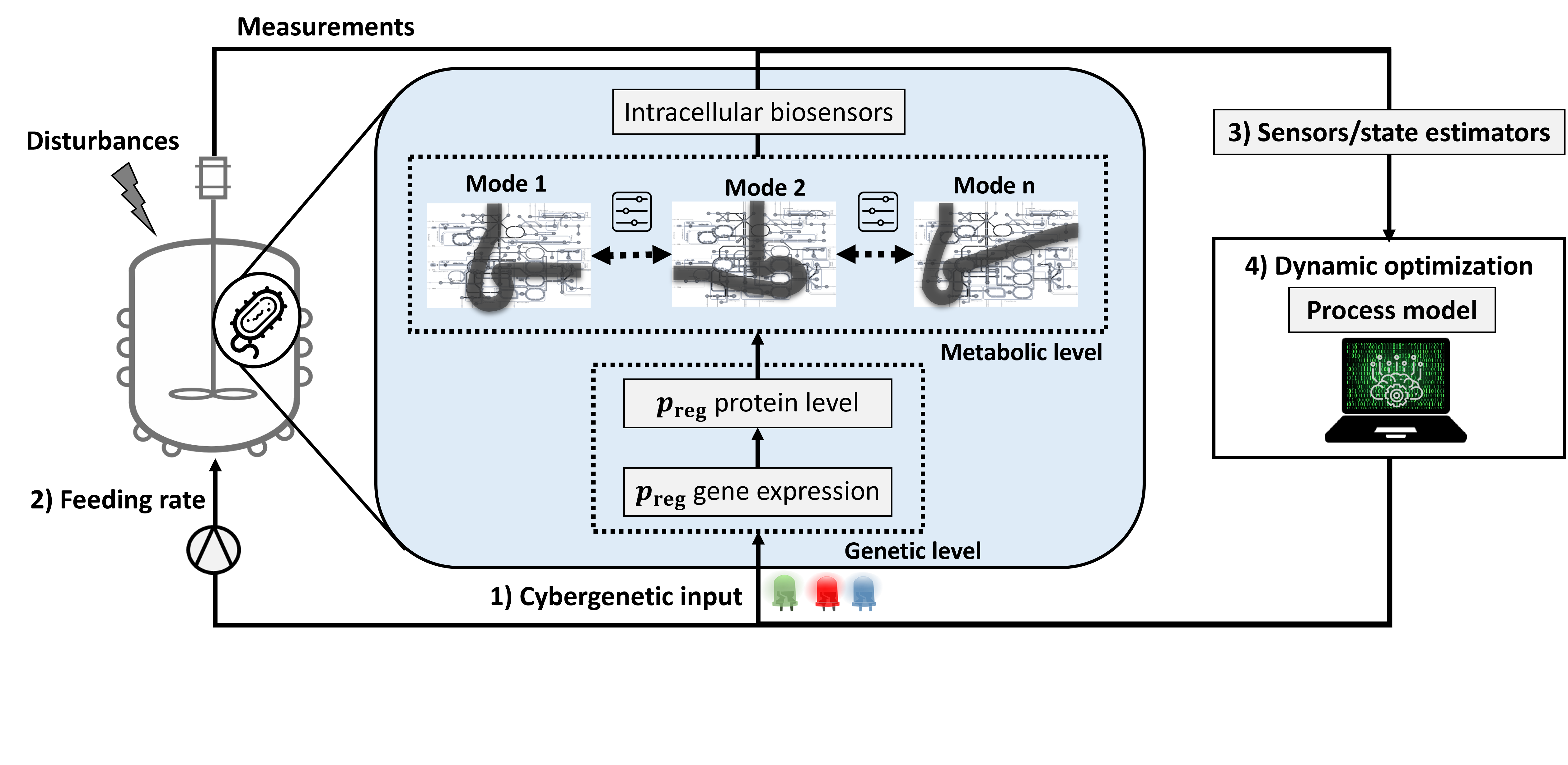}
\caption{Overview of metabolic cybergenetics in fed-batch regime. Key metabolism-relevant proteins such as enzymes $p_{\text{reg}}\in \mathbb{R}^{n_{p_{\text{reg}}}}$ are under the regulation of inducible gene expression systems to enable different metabolic modes over time using external inputs (e.g., light intensity). Model-based optimization finds the optimal inputs to the plant. The process outcome can be monitored via (bio)sensors and state estimators. Repeated solution of the optimization leads to feedback control.}
\label{fig:overview}
\end{center}
\end{figure*}

A reasonably good model capable of relating inducible gene expression to changes in the metabolic flux distribution and potential resource burden thus becomes fundamental for advancing in our quest. Unfortunately, so far only very simple models have been used in the context of cybergenetics, often based on phenomenological relations (cf. e.g., \cite{milias-argeitis_automated_2016,lovelett_dynamical_2021,gutierrez_mena_dynamic_2022}). In our opinion, these models do not allow capturing all the important phenomena required for model-based control of \textit{metabolic} cybergenetic systems.

Thus, as the core contribution of this work, we propose a modeling framework for metabolic cybergenetics, which is combined with model-based optimization and predictive control to dynamically modulate intracellular metabolic fluxes for bioprocess optimization. Without loss of generality, we focus on fed-batch processes due to their advantages compared to pure batch setups. Fed-batch processes include a concentrated feed that supplies fresh medium to the bioreactor, thus extending the production phase. This provides additional dynamic inputs (feed rates) to the system, and allows for higher productivity and more concentrated product streams. It furthermore provides an efficient way to handle processes with substrate inhibition \citep{doran_bioprocess_2013,liu_bioprocess_2020}. The proposed fed-batch metabolic cybergenetic platform comprises four major components (see Fig. \ref{fig:overview}): 1) a cybergenetic input capable of inducing gene expression dynamically, 2) a manipulatable substrate feeding stream, 3) online (bio)sensors and state estimators that monitor and estimate the state of the process, 4) and model-based optimization that operates in a closed-loop and fully-automated fashion.

The remainder of the paper is structured as follows. In Section \ref{sec:modelling_framework} we outline a dynamic constraint-based cybergenetic modeling approach that integrates metabolism, resource allocation and inducible gene expression. The derived model is used to support model-based optimization, feedback control and state estimation of metabolic cybergenetics (Sections \ref{sec:optimal_control} and \ref{sec:MPC_formulation}). In Section \ref{sec:case_study} we evaluate our framework considering the optogenetic modulation of the ATPase F\textsubscript{1}-subunit\footnote{From now on, we will refer to the F\textsubscript{1}-subunit of the ATPase enzyme/gene as the \say{ATPase enzyme/gene}.} in the anaerobic lactate fermentation by \textit{Escherichia coli} for improved yield and productivity. We consider a \textit{fed-batch} regime with \textit{non-homogeneous} light penetration, applying both open-loop \textit{and} feedback control. Note that in a previous study we covered only open-loop optimization in classic batch processes with homogeneous light penetration \citep{espinel_opt_2022}.

\section{Modeling for dynamic optimization and control of fed-batch metabolic cybergenetics}
\label{sec:modelling_framework}
We use an extended constraint-based modeling approach for capturing the combined dynamics of metabolism, resource allocation and inducible gene expression. Constraint-based models are usually underdetermined \citep{gottstein_constraint-based_2016,klamt_analyzing_2022}. Therefore, they are often formulated as optimization problems with biologically sound objective functions and subject to constraints. Constraint-based models generally require a smaller number of parameters (e.g., flux bounds) and can be structurally simpler compared to kinetic models \citep{saa_formulation_2017,yasemi_modelling_2021}.

Without loss of generality, we consider that cells are composed of metabolic enzymes, ribosomes and quota elements\footnote{Quota elements include, e.g., DNA, lipids, carbohydrates, non-catalytic proteins plus other small molecules.}. These biomass components are contained in the molar vector $\bm{p} \in \mathbb{R}^{n_p}$ such that the amount of biomass $B\in \mathbb{R}$ can be computed as
\begin{equation} \label{eq:biomass}
{B(t)=v_L(t)(\bm{b}^T\bm{p}(t))},
\end{equation}
where $t$ is the time, $v_L\in \mathbb{R}$ is the bioreactor volume and $\bm{b} \in \mathbb{R}^{n_p}$ is a vector of the corresponding molecular weights of $\bm{p}$. Therefore, $\bm{b}^T\bm{p}(t)$ is a scalar quantity corresponding to the biomass concentration in $\mathrm{g}/\mathrm{L}$. Remark that we use bold fonts for vectors and matrices, and non-bold fonts for scalar variables and parameters.

We collect in $\bm{p}$ both the concentrations of regulated proteins $\bm{p_{\mathrm{reg}}}\in \mathbb{R}^{n_{p_{\mathrm{reg}}}}$ and the concentrations of the remaining unregulated biomass components $\bm{p_{\mathrm{unr}}}\in \mathbb{R}^{n_{p_{\mathrm{unr}}}}$, such that $\bm{p}(t)=\begin{bmatrix}\bm{p_{\mathrm{reg}}}(t)^\tran, & \bm{p_{\mathrm{unr}}}(t)^\tran\end{bmatrix}^\tran$. In this text, the term \say{regulated} refers to the fact that the protein expression is under cybergenetic control, externally modulated via a suitable genetic system such as a light-inducible gene expression system \citep{liu_programming_2018,lindner_optogenetics_2021}.
Note that the regulated proteins can comprise enzymes directly involved in metabolic pathways, which is the main focus of this paper, but can in principle also include (anti)toxin proteins or antibiotic-resistance conferring proteins for growth modulation. The \say{unregulated} biomass components typically contain, e.g., the metabolic enzymes that are \textit{not} under cybergenetic control, ribosomes and quota elements. Hereafter, we will omit writing the dependency of the variables with respect to time when clear from the context.

With regards to the tunable gene expression systems, we differentiate between the inputs manipulated by the controller $\bm{u_s} \in \mathbb{R}^{n_{u}}$ (e.g., light intensity) and the values perceived by the cells inside the bioreactor $\bm{u_c} \in \mathbb{R}^{n_{u}}$. The distinction is necessary, as depending on the characteristics of the input and bioreactor, $\bm{u_s}=\bm{u_c}$ might not hold. This is especially relevant in large-scale setups where conditions tend to be less homogeneous, or where the input values received by the cells might depend on the cell density, e.g., due to turbidity.

In Section \ref{sec:case_study} we will derive $\bm{u_c}$ to account for light penetration in the context of optogenetics. For the sake of generality, we consider that the input perceived by the cells is given by a function $\bm{f_{u}}:\mathbb{R}^{n_{u}} \times \mathbb{R}^{n_{x}} \times \mathbb{R}^{n_{\theta_u}}\rightarrow  \mathbb{R}^{n_u}$ which maps the input at the source $\bm{u_s}$ to an \emph{average} input $\bm{\bar{u}_{c}}$. Hence,
\begin{equation} \label{eq:u_average}
\bm{\bar u_{c}}=\bm{f_{u}}(\bm{u_{s}},\bm{x},\bm{\theta_{u}}),
\end{equation}
where $\bm{x} \in \mathbb{R}^{n_{x}}$ can include in principle all the model states, $\bm{\theta_{u}} \in \mathbb{R}^{n_{\theta_u}}$ comprises possible parameters of $\bm{f_{u}}(\cdot)$ and $\bm{\bar u_{c}}$ is the average value of $\bm{u_{c}}$ in the bioreactor following well-mixed conditions. Introducing $\bm{\bar u_{c}}$ simplifies the model as it is limited to changes in time and not in space, for example, while still accounting for average input gradients. 

We describe the resulting change in the amount of regulated protein as
\begin{equation} \label{eq:gene_expression_fed_batch}
\begin{aligned}
&\odv{\left(v_L\bm{p_{\mathrm{reg}}} \right)}{t} = \bm{F_{\mathrm{reg}}}(B,\bm{\bar u_{c}})-\bm{D_{\mathrm{reg}}}(\bm{p_{\mathrm{reg}}}),
\end{aligned}
\end{equation}
where $\bm{F_{\mathrm{reg}}}: \mathbb{R} \times  \mathbb{R}^{n_u}  \rightarrow \mathbb{R}^{n_{p_{\mathrm{reg}}}}$ and $\bm{D_{\mathrm{reg}}}: \mathbb{R}^{n_{p_\mathrm{{reg}}}} \rightarrow \mathbb{R}^{n_{p_{\mathrm{reg}}}}$ are functions representing protein production and degradation, respectively.

Cells possess transcription factors that can switch between active and inactive states at a rate dictated by a specific signal. When active transcription factors bind the promoter region of a regulated gene, they can activate or repress the transcription process. Ribosomes catalyze the translation of the messenger ribonucleic acid, resulting from the transcription process, into proteins. In bacteria such as \emph{E. coli}, transcription and translation are highly coupled, meaning that translation occurs at the same time as active transcription \citep{yang_transcription_2019,scull_transcriptional_2021}. Therefore, we propose to combine these two processes into lumped dose-response functions $\bm{\eta}: \mathbb{R}^{n_{u}} \rightarrow \mathbb{R}^{n_{p_{\mathrm{reg}}}}$, hence
\begin{equation} \label{eq:F_reg_fed_batch}
{\bm{F_{\mathrm{reg}}}(B,\bm{\bar u_{c}}) = B\bm{\eta}(\bm{\bar u_{c}})}.
\end{equation} 

For the degradation of regulated proteins, we consider both the effect of cell dilution due to growth and intrinsic protein turnover. The latter is captured by $\bm{D_{\mathrm{reg}}}$. On the other hand, the dilution of the regulated proteins is implicitly considered in our modeling framework. That is, the proportion of regulated proteins within the cell is \textit{diluted} by the production of the remaining biomass components.

We connect the dynamics of the regulated proteins to the overall metabolism and cell resource allocation via dynamic enzyme-cost flux balance analysis (deFBA), a constraint-based metabolic framework that considers resource allocation constraints \citep{waldherr_dynamic_2015,jabarivelisdeh_adaptive_2020,jabarivelisdeh_optimization_2018}.

The amount of extracellular metabolites is modeled as
\begin{equation} \label{eq:dynamics_z_fed_batch}
\begin{aligned}
&\odv{\left(v_L\bm{z}\right)}{t} = F_{\mathrm{in}}\bm{z_{\mathrm{in}}} + v_L(\bm{S_z}\bm{V}) - \bm{D_z}(\bm{z}),
\end{aligned}
\end{equation}
where $\bm{z} \in \mathbb{R}^{n_z}$ is the molar vector of extracellular metabolites, $\bm{S_z} \in \mathbb{R}^{n_z\times n_V}$ is the stoichiometric matrix of $\bm{z}$ and $\bm{V} \in \mathbb{R}^{n_V}$ includes the fluxes for transport, metabolic and biomass-producing reactions in molar amount per time. $\bm{D_{z}}: \mathbb{R}^{n_z} \rightarrow \mathbb{R}^{n_z}$ captures the degradation of $\bm{z}$, $F_{\mathrm{in}} \in \mathbb{R}$ is the feeding rate and $\bm{z_{\mathrm{in}}} \in \mathbb{R}^{n_z}$ comprises the feed concentrations of $\bm{z}$. In the model we assume that the feed contains only substrates.

The cell needs to invest resources to manufacture its components. Thus, changing the expression of a regulated protein is expected to influence the production rate of other biomass components and the resulting metabolic flux distribution since resources are limited and shared within the cell. Note that including biomass-producing reactions in the network is a way to capture the resource \textit{cost} because we explicitly consider the required stoichiometric precursor and energy equivalents for the synthesis of all biomass components. With this in mind, the amount of unregulated biomass components follows
\begin{equation} \label{eq:dynamics_p_unr_fed_batch}
\begin{aligned}
&\odv{\left(v_L\bm{p_{\mathrm{unr}}}\right)}{t} = v_L(\bm{S_{p_{\mathrm{unr}}}}\bm{V}) - \bm{D_{\mathrm{unr}}}(\bm{p_{\mathrm{unr}}}),  
\end{aligned}
\end{equation}
where $\bm{S_{p_{\mathrm{unr}}}} \in \mathbb{R}^{n_{p_{\mathrm{unr}}}\times n_V}$ is the stoichiometric matrix of $\bm{p_{\mathrm{unr}}}$ and $\bm{D_{\mathrm{unr}}}: \mathbb{R}^{n_{p_\mathrm{{unr}}}} \rightarrow \mathbb{R}^{n_{p_{\mathrm{unr}}}}$ describes the degradation of $\bm{p_{\mathrm{unr}}}$. For the regulated proteins, we add the following constraint
\begin{equation} \label{eq:cost_p_reg}
{\bm{V_{p_{\mathrm{reg}}}}-\odv{\bm{p_{\mathrm{reg}}}}{t}=\textbf{0}},
\end{equation}
where $\bm{V_{p_{\mathrm{reg}}}}\in \mathbb{R}^{n_{p_{\mathrm{reg}}}}$ contains the corresponding regulated protein-producing reaction fluxes.

We consider quasi-steady-state dynamics for the amount of intracellular metabolites
\begin{equation} \label{eq:dynamics_m}
{\odv{\left(v_L\bm{m}\right)}{t}=v_L(\bm{S_m}\bm{{V}})=\textbf{0}},
\end{equation}
where $\bm{m} \in \mathbb{R}^{n_m}$ is the molar vector of intracellular metabolites and $\bm{S_m} \in \mathbb{R}^{n_m\times n_V}
$ is the stoichiometric matrix of $\bm{m}$.

The metabolic fluxes of reactions catalyzed by enzymes in $\bm{p_{\mathrm{unr}}}$ are constrained by the corresponding catalytic enzyme concentration and catalytic constant ($\bm{k_{\mathrm{cat}}} \in \mathbb{R}^{n_{\mathrm{cat}}}$)
\begin{equation} \label{eq:k_cat_p_unr}
{\sum \limits_{j\in \mathrm{cat}_\mathrm{unr}} \left | \frac{V_j}{k_{\mathrm{unr},j}} \right |\leq p_{\mathrm{unr}_i}, \quad  \forall i \in [1,n_{p_{\mathrm{unr}}}]},
\end{equation}
where $\mathrm{cat}_\mathrm{unr}(i)$ are all the reactions catalyzed by an enzyme $p_{\mathrm{unr}_i}$ and $| \cdot |$ refers to the absolute value operator.

The metabolic fluxes of reactions catalyzed by enzymes in $\bm{p_{\mathrm{reg}}}$ are constrained by
\begin{equation} \label{eq:k_cat_p_reg}
{\sum \limits_{j\in \mathrm{cat}_\mathrm{reg}(i)} \left | \frac{V_j}{k_{\mathrm{reg},j}} \right | = p_{\mathrm{reg}_i}, \quad \forall i \in [1,n_{p_{\mathrm{reg}}}]},
\end{equation}
where $\mathrm{cat}_\mathrm{reg}(i)$ is the set of reactions catalyzed by $p_{\mathrm{reg}_i}$. 

We consider enzyme saturation conditions. Thus, Eq. \eqref{eq:k_cat_p_unr} takes the product of the enzyme concentrations and the catalytic constants as upper bounds for the metabolic fluxes. While Eq. \eqref{eq:k_cat_p_unr} is an inequality constraint, we use an equality constraint in Eq. \eqref{eq:k_cat_p_reg} under the assumption that we have control over the fluxes catalyzed by $p_{\mathrm{reg}_i}$. In other words, we \textit{shift} this degree of freedom from the cell to an external controller.

A fraction $\varphi_Q\in[0,1]$ of the biomass dry weight corresponds to a lumped quota compound $p_Q \in \mathbb{R}$
\begin{equation} \label{eq:quota}
{\varphi_Q(\bm{b}^T\bm{p})\le p_Q, \, p_Q \in \bm{p_{\mathrm{unr}}}}.
\end{equation}

The bioreactor liquid volume changes over time due to the substrate feeding rate $F_{\mathrm{in}}$
\begin{equation} \label{eq:volume_fed_batch}
\begin{aligned}
&\odv{v_L}{t} = F_{\mathrm{in}}.
\end{aligned}
\end{equation}

Metabolic fluxes are constrained by biologically feasible lower and upper bounds
\begin{equation} \label{eq:V_min_max}
{\bm{V_{\mathrm{min}}} \le \bm{V}\le \bm{V_{\mathrm{max}}}}.
\end{equation}

Similarly, we consider feasible bounds for the dynamic states
\begin{equation} \label{eq:x_box}
{\bm{p_{\mathrm{min}}} \le \bm{p}\le \bm{p_{\mathrm{max}}}}, \, 
{\bm{z_{\mathrm{min}}} \le \bm{z}\le \bm{z_{\mathrm{max}}}}, \, 
{v_{L_{\mathrm{min}}}} \le {v_L}\le {v_{L_{\mathrm{max}}}}
.
\end{equation}

The conditions of the system at the initial process time $t_0$ are
\begin{equation} \label{eq:initial_conditions}
\begin{aligned}
\bm{p}(t_0)=\bm{p_{0}}, \, \bm{z}(t_0)=\bm{z_0}, \, v_L(t_0)=v_{L0}.
\end{aligned}
\end{equation}

Summarizing, we express the resulting dynamic constraint-based model for fed-batch metabolic cybergenetic systems in terms of the following general dynamic optimization problem\footnote{The model allows describing batch systems by setting $\dot{v}_L=0$. For continuous processes, one can include an additional flow rate leaving the bioreactor.}
\begin{maxi!}<b>
    {\bm{V}(\cdot)}{\int_{t_0}^{t_0 + \Delta t_{\mathrm{bio}}}F_{\mathrm{V}}(\cdot)\diff t,}{\label{eq:model_fed_batch}}{} 
    \addConstraint{\textrm{Eqs.}\:(\ref{eq:biomass})-\:(\ref{eq:initial_conditions}),}{}{}
\end{maxi!}
where $F_{\mathrm{V}}(\cdot)$ is the objective function that the cell \textit{optimizes}, usually one assumes it is the maximization of cell growth, and $V(\cdot)$ is a function of the resulting metabolic flux distribution. Solving this dynamic optimization problem allows one to simulate and predict the
cell´s behavior, as demonstrated in Section \ref{sec:case_study}. We will use this dynamic constraint-based model as a basis for optimizing and controlling the process.

\section{Optimal control for metabolic cybergenetics}
\label{sec:optimal_control}
Based on the derived dynamic constraint-based model, we aim to find the optimal input trajectories to drive the cell metabolism towards maximizing a desired performance criterion described by a cost function $J(\cdot)$. Let us collect all the process inputs (manipulated variables) in $\bm{u_p}$, i.e.,  $\bm{u_p} := [\bm{u_s}(\cdot)^\tran, F_{\mathrm{in}}(\cdot)]^\tran$, and all the model parameters in the vector $\bm{\theta}$. Recall that $\bm{x}$ contains all the dynamic states, i.e., $\bm{x}:=[\bm{p_{\mathrm{reg}}}^\tran,\bm{z}^\tran,\bm{p_{\mathrm{unr}}}^\tran,v_L]^{\tran}$. To find the optimal inputs to the plant, we formulate an optimal control problem
\begin{maxi!}<b>
    {\bm{u_p}}{J(\cdot),}{\label{eq:op_con_fed_batch}}{}
    \addConstraint{(\ref{eq:model_fed_batch})}{}{}, 
    \addConstraint{\bm{0} \leq \bm{g}(\bm{x},\bm{u_p},\bm{\theta})}{}{}.\label{eq:other_cons}
\end{maxi!}

Solving \eqref{eq:op_con_fed_batch} is a bilevel optimal control problem as the dynamic constraint-based model in \eqref{eq:model_fed_batch} involves an optimization on its own. Eq. \eqref{eq:other_cons} captures additional state and input constraints. 

$J(\cdot)$ can be defined in several ways based on specific goals. One may want to maximize production, maintain a desired set-point, and follow a reference trajectory, among other possibilities. Eq. \eqref{eq:other_cons} can include, e.g., physical-, safety- or economic-related process constraints. If the process is run in batch mode, then $\bm{u_s}(\cdot)$ can be set as the only optimization degree of freedom. The optimal control problem in \eqref{eq:op_con_fed_batch} is an open-loop optimization, as only the initial conditions of the process states are used to compute an optimal input trajectory which is then applied to the plant without feedback. Doing so, would not allow reacting to unknown disturbances or model-plant mismatch.  

\section{Model predictive control for metabolic cybergenetics}
\label{sec:MPC_formulation}
As we consider fed-batch processes, we use shrinking horizon model predictive control (MPC) to mitigate the effects of process uncertainty such as model-plant mismatch and disturbances \citep{rawlings_model_2020,findeisen_introduction_2002}, i.e., to mitigate the challenges of open-loop control.

\subsection{Shrinking horizon model predictive control}
 In MPC, the optimal control problem is evaluated repetitively at given sampling times. At these sampling instances the states of the system are measured or estimated with an observer. This introduces feedback since the information on the current system states is passed to the controller and corrective control actions can be taken. Let $t_k$ be the sampling times at which measurements are taken. Without loss of generality, we assume that state measurements are available at equidistant sampling times, i.e., $t_k :=kh_s$ where $k \in \mathbb{N}_0$ and $h_s$ is a fixed sampling interval. Furthermore, we assume that the controller predicts up to the final time $t_f:= N h_s$, where $N \in \mathbb{N}$ is the number of steps in the horizon. Therefore, the prediction horizon \textit{shrinks} at every sampling time. The shrinking horizon MPC at time $t_k$ reads
\begin{mini!}<b>
    {\bm{u_p}}{J(\cdot),}{\label{eq:optimal_mpc}}{\label{eq:MPC_obj}}
    \addConstraint{\max_{\bm{V}(\cdot)}\int_{t_k}^{t_k + \Delta t_{\mathrm{bio}}} F_{\mathrm{V}}(\cdot)\diff t, \label{eq:obj_mpc_defba}}{}{}
    \addConstraint{\text{s.t.}\quad \textrm{Eqs.\:(\ref{eq:biomass})-\:(\ref{eq:x_box}),}}{}{} 
    \addConstraint{\phantom{\text{s.t.}} \quad \bm{x}(t_k)=\bm{\tilde{x}_k}\label{eq:mpc_x0},}{}{}
    \addConstraint{\textrm{Eq.\:(\ref{eq:other_cons}), }}{}{}
\end{mini!}
where $t \in[t_k, t_f]$ and $\bm{\tilde{x}_k}$ indicates the measured value of $\bm{x}$. 

We assume that the culture volume can be monitored straightforwardly based on the applied feeding rate, and there is a range of online sensors available for the extracellular metabolite concentrations \citep{reardon_practical_2021,reyes_modern_2022,fung_shek_taking_2021}. Therefore, monitoring $v_L$ and $\bm{z}$ is technically possible with the present technologies. However, typically there are no commercial sensors for the complete intracellular biomass composition. To circumvent this challenge, some state estimators have been proposed for \textit{reconstructing} the biomass composition \citep{jabarivelisdeh_adaptive_2020,espinel_fie_2022}. In the next section, we briefly describe the use of a full information estimator --an optimization-based estimator that considers the process dynamics and process constraints, as well as past and current measurements. For more details, we refer the reader to \cite{espinel_fie_2022}.

\begin{figure*}[htb]
\begin{center}
\includegraphics[scale=0.45]{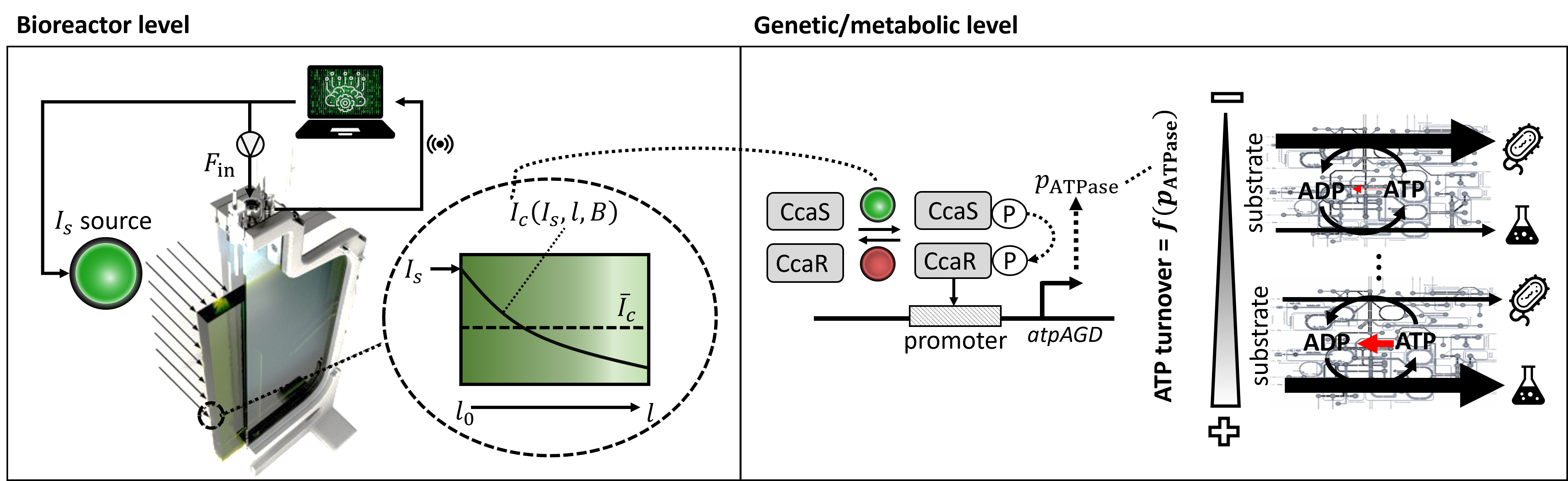}\caption{Overview of the considered example using the Ccas/CcaR optogenetic system for modulating the expression of the ATPase enzyme. The left side shows the considered flat-panel (photo)bioreactor and the metabolic cybergenetic control scheme, including the average light intensity inside the culture. On the right side, we show the effect of different expression levels of ATPase on the cell´s metabolism. Picture of the bioreactor adapted from \cite{pfaffinger_model-based_2016}.} \label{fig:case_overview}
\end{center}
\end{figure*}

\subsection{Reconstructing unmeasured cell components}
\label{sec:FIE}
Let $(\cdot)_i$ be a general optimization variable calculated at time $t_i$. We collect the dynamic equations \eqref{eq:gene_expression_fed_batch}, \eqref{eq:dynamics_z_fed_batch},  \eqref{eq:dynamics_p_unr_fed_batch} and \eqref{eq:volume_fed_batch} in the vector function $\bm{f}(\bm{x},\bm{u_p},\bm{\theta})$. As with the MPC, we assume for simplicity of presentation equidistant sampling times for the estimator, although non-equidistant sampling times are also possible. At time $t_k$, a full information estimator can be formulated by solving the following optimization problem
\begin{mini!}<b>
    {\bm{x_{0}},  \bm{\theta} , \bm{w}}{\left\Vert \begin{bmatrix}\bm{x_0}  \\ \bm{\theta} \end{bmatrix} - \begin{bmatrix}\bm{\hat{x}_0} \\ \bm{\hat{\theta}}\end{bmatrix} \right\Vert_\mathbf{P}^2 + \sum_{i=0}^{k} \Vert \bm{y}(t_i) - \bm{\tilde{y}_i}\Vert_\mathbf{R}^2 +}{\label{eq:optimal_mhe}}{\label{eq:obj_mhe}}
    \breakObjective{+ \Vert \bm{w_i} \Vert_\mathbf{Q}^2} \nonumber
    \addConstraint{\max_{\bm{V}(\cdot)}\int_{t_i}^{t_i+\Delta t_{\mathrm{bio}}} F_{\mathrm{V}}(\cdot)\diff t \label{eq:obj_mhe_dfba}}{}{}
    \addConstraint{\text{s.t.}\quad \bm{x}(t_i + h_s)= \bm{x}(t_i)^\tran +\label{eq:x_dynamics_mhe}}{}{} 
    \addConstraint{\phantom{\text{s.t.}}\quad \int_{t_i}^{t_i+h_s}\bm{f}(\bm{x},\bm{u_p},\bm{\theta}) \diff t  + \bm{w_i}, \nonumber }{}{}
    \addConstraint{\phantom{\text{s.t.}} \quad \bm{0} \leq \bm{c}(\bm{x}(t_i),\bm{u_p}(t_i),\bm{\theta}),}{}{}
    \addConstraint{\phantom{\text{s.t.}} \quad \bm{y}(t_i) = \bm{h}(\bm{x}(t_i),\bm{u_p}(t_i),\bm{\theta}),}{}{}
    \addConstraint{\bm{0} \leq \bm{g}(\bm{x}(t_i),\bm{u_p}(t_i),\bm{\theta}),}{}{}
\end{mini!}
where $\Vert \bm{a} \Vert_\mathbf{A}^2 := \bm{a}^\tran \mathbf{A} \bm{a}$ and $i \in [0,...,k]$. $\mathbf{P}$, $\mathbf{R}$ and $\mathbf{Q}$ are weighting matrices of appropriate dimensions, $\bm{c}: \mathbb{R}^{n_x} \times \mathbb{R}^{n_u+1} \times \mathbb{R}^{n_{\theta}} \rightarrow \mathbb{R}^{n_c}$
are the model constraints and $\bm{h}: \mathbb{R}^{n_x} \times \mathbb{R}^{n_u+1} \times \mathbb{R}^{n_{\theta}} \rightarrow \mathbb{R}^{n_y}$ are the measurement equations. The optimization variables are the initial condition $\bm{x_0}$, the parameter $\bm{\theta}$ and the state noise $\bm{w}:=[\bm{w_0}^\tran,...,\bm{w_{k}}^\tran]^\tran$. We indicate with $(\cdot)^*$ the solution of the full information estimation problem and with $(\hat{\cdot})$ the \emph{prior information} of a variable. With $\bm{x_0}^*$, $\bm{w}^*$ and $\bm{\theta}^*$ we reconstruct the states at $t_k$ which can be used in the MPC. The full information estimator considers \emph{all} the measurements; instead, if only the measurements in a given time window are used, one refers to a moving horizon estimator \citep{rawlings_model_2020,elsheikh_comparative_2021}.

\begin{figure} [htb]
\begin{center}
\includegraphics[scale=0.40]{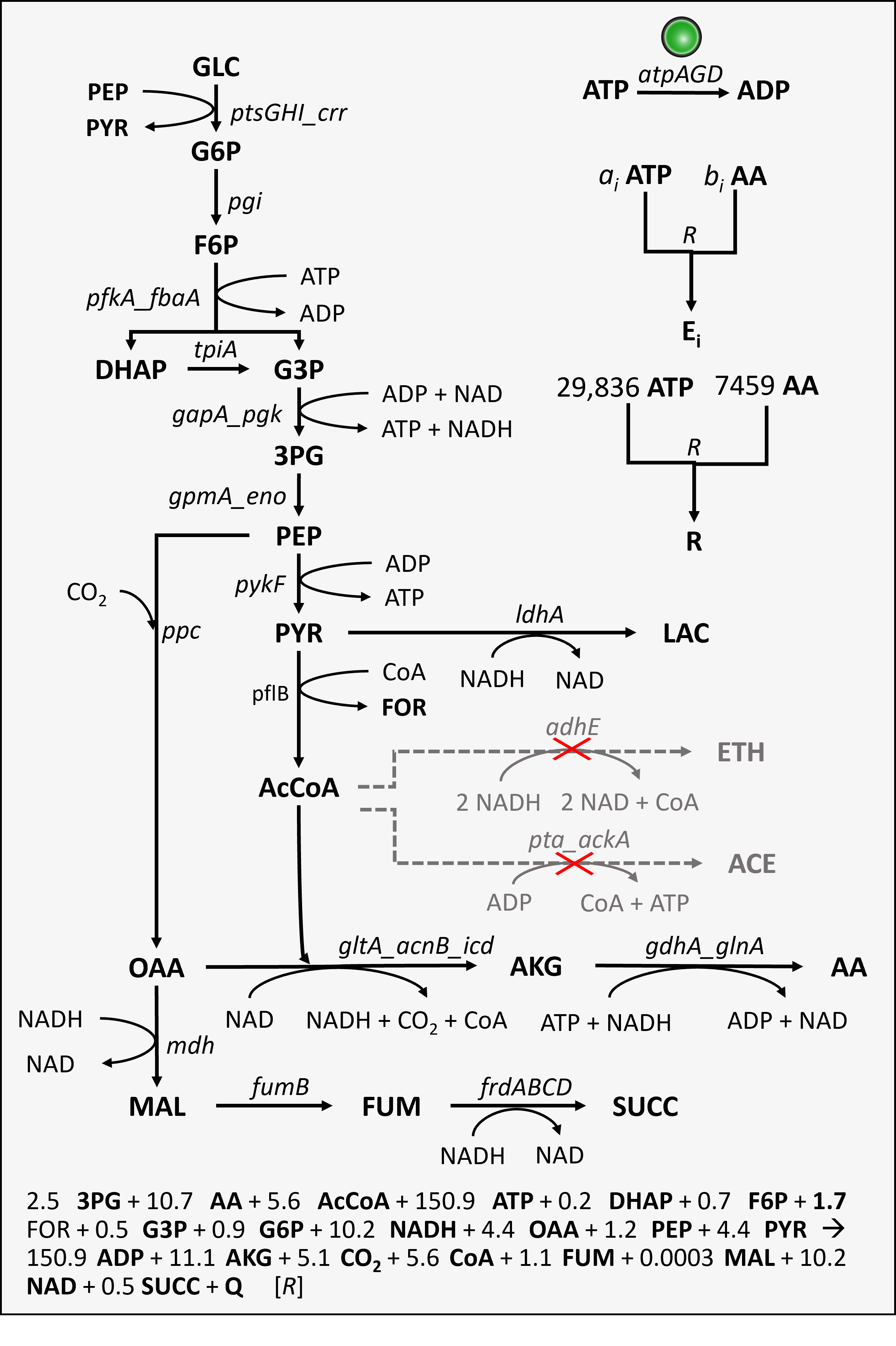}
\\[-0.2cm]
\fbox{\begin{minipage}{24em}
\small{3PG: 3-phospho-D-glycerate; AA: amino acid; AcCoA: acetyl-CoA; ACE: acetate; ADP: adenosine diphosphate; AKG: alpha-ketoglutarate; ATP: adenosine triphosphate; CO\textsubscript{2}: carbon dioxide; CoA: coenzyme A; DHAP: dihydroxyacetone phosphate; ETH: ethanol; E\textsubscript{i}: enzyme i; F6P: fructose 6-phosphate; FOR: formate; FUM: fumarate; G3P: glyceraldehyde 3-phosphate; G6P: glucose 6-phosphate; GLC: glucose; LAC: lactate; MAL: malate; NAD: nicotinamide adenine dinucleotide; NADH: NAD-reduced; OAA: oxaloacetic acid; PEP: phosphoenolpyruvate; PYR: pyruvate; Q, quota; R, ribosome; SUCC: succinate.}
\end{minipage}}
\caption{Scheme of the resource allocation model for the anaerobic lactate fermentation by \textit{E. coli} KBM10111. Catalytic species are shown in italics (e.g., \textit{atpAGD} refers to the genes of the ATPase enzyme). Some enzymes are lumped via underscore symbols. In gray we depict the blocked pathways. Adapted from \cite{espinel_ATP_w_2022}.} \label{fig:pathway}
\end{center}
\end{figure}

\section{Example: optogenetic control of ATPase for anaerobic lactate fermentation by E.coli}
\label{sec:case_study}
 We consider the anaerobic lactate fermentation by \textit{E. coli} using glucose as substrate, with optogenetic control of the ATPase enzyme complex, c.f. Fig. \ref{fig:case_overview}. We only have one regulated protein, hence, $\bm{p_{\mathbf{reg}}} := p_{\mathrm{ATPase}}$. The latter enzyme is responsible for catalyzing the hydrolysis reaction of ATP into ADP. We focus on the \textit{E. coli} KBM10111 strain, engineered with gene deletions of \textit{adhE} (aldehyde-alcohol dehydrogenase), \textit{ackA} (acetate kinase) and \textit{pta} (phosphate acetyltransferase) \citep{hadicke_enforced_2015}. Under these conditions, lactate synthesis from pyruvate is required to balance the redox cofactors generated during glycolysis. Since glycolysis renders net ATP gain, lactate production is linked to ATP synthesis (see Fig. \ref{fig:pathway}). In such cases, where the product pathway is linked to net ATP formation, it has been shown that an \textit{enforced} ATP turnover or \textit{wasting} can lead to an increase in the substrate uptake and the metabolic flux through the ATP-producing pathway as a way to counterbalance the ATP loss (cf. e.g. \cite{hadicke_enforced_2015,zahoor_atpase-based_2020,boecker_increasing_2021}). Dynamic manipulation of the ATPase expression, and thereby the ATPase flux, can thus be exploited to modulate the product yield and volumetric productivity in bioprocesses \citep{espinel_ATP_w_2022,espinel_opt_2022}.

We consider the Ccas/CcaR optogenetic system \citep{olson_characterizing_2014} for modulating the ATPase expression. CcaS is a sensor histidine kinase that is activated with green light ($\lambda _{535 nm}$). Active CcaS phosphorylates the cognate response regulator CcaR. Phosphorylated (active) CcaR enables the transcription of the target genes. In contrast, CcaS is inactivated with red light ($\lambda _{650 nm}$), thereby blocking transcription. From now on, let $\bm{u_s} := I_s$ and $\bm{\bar u_c} := \bar I_c$, where $I_s$ is the green light intensity manipulated by the controller and $\bar I_c$ is the average of $I_c$, i.e., of the green light intensity perceived by the cells inside the bioreactor. Therefore, the process inputs comprise one cybergenetic input plus the substrate feeding rate. 

Furthermore, we consider a flat-panel photobioreactor, consisting of two flat surfaces joint by a thin gap, thereby creating a rectangular channel \citep{chanquia_photobioreactors_2021}. The bioreactor is illuminated from one side by a green light source. This geometry is known to maximize the illumination area per culture volume, hence it is appealing for optogenetics.

\subsection{Model and process considerations}
We assume that the dose-response function for ATPase expression follows a Hill function \citep{olson_characterizing_2014}
\begin{equation} \label{eq:hill_activation}
\begin{aligned}
\eta_{\mathrm{ATPase}}\left(\bar I_{c}\right)=\alpha+\beta \frac{\bar I_{c}^\delta}{K^\delta+\bar I_{c}^{\delta}},
\end{aligned}
\end{equation}
where $\alpha$ is an input-independent basal rate of production (e.g., due to promoter \textit{leakage} or constitutive expression), $\beta$ is an input-dependent maximum rate of production, $K$ is a saturation constant and $\delta$ is the Hill coefficient.

We assume that light penetration inside the bioreactor is not homogeneous as the cells interfere with the light beam. Let $l$ be the length between the two plates of the bioreactor. We set up a balance over an infinitesimally small distance $\diff l$, assuming that the light hits perpendicularly with respect to the illuminated flat surface and that the culture is well-mixed. After integrating from $l_0$ to $l$ we obtain
\begin{equation} \label{eq:integral_light}
\begin{aligned}
&I_c(l,B) = I_c(l_0,B)e^{-a_{\lambda}B(l-l_0)},
\end{aligned}
\end{equation}
where $I_c(l_0,B)=I_s$, and $a_\lambda$ is a lumped biomass-specific constant that accounts for light scattering and absorption effects. Note that the latter equation follows a similar derivation as the Lambert-Beer law \citep{hofmann_simon_grkovic_jones_2014}. We obtain $\bar{I}_c$ from the mean integral of $I_c(l,B)$ from 0 to $l$
\begin{equation} \label{eq:average_light}
\begin{aligned}
&\bar{I}_c (B)= \frac{I_s}{a_{\lambda}Bl}\left(1-e^{-a_{\lambda}Bl}\right).
\end{aligned}
\end{equation}

The dynamics of the lactate fermentation are based on an existing deFBA model \citep{espinel_ATP_w_2022}. See Fig. \ref{fig:pathway} for a summary of the resource allocation model. In general, the model contains 34 fluxes: 16 metabolic reactions and 18 biomass-producing reactions. From the latter, 16 reactions are for the synthesis of catalytic enzymes, one for ribosomes and another one for a lumped quota compound. It considers 5 species in $\bm{z}$ (glucose, lactate, formate, succinate and carbon dioxide), 18 species in $\bm{m}$ and 18 species in $\bm{p}$. The cell composition (g/g biomass) is 0.06 catalytic enzymes, 0.38 non-catalytic enzymes, 0.27 ribosomes and 0.29 other components (DNA, lipids, carbohydrates, etc.), hence $\varphi_Q = 0.38 + 0.29$. The model already includes an ATPase-producing reaction. The \textit{cost} of the biomass-producing reactions are expressed in terms of amino acids and ATP equivalents. The model parameters include 34 catalytic constants and 18 molecular weights. Furthermore, the model takes $F_{\mathrm{V}} = B(t)$ in \eqref{eq:model_fed_batch}.  

Additionally, we consider negligible degradation in Eqs. \eqref{eq:dynamics_z_fed_batch}-\eqref{eq:dynamics_p_unr_fed_batch}, while for the ATPase enzyme
\begin{equation} \label{eq:D_reg_fed_batch}
{D_{\mathrm{ATPase}}= d_{\mathrm{ATPase}}p_{\mathrm{ATPase}}v_L},
\end{equation} 
where $d_{\mathrm{ATPase}}$ is a constant ATPase degradation rate.

In Table \ref{tb:optogenetic_parameters} we summarize the model parameters for the CcaS/CcaR module and flat-panel bioreactor, as well as the process initial conditions. The cost function is chosen to maximize the lactate concentration at the end time of the process, i.e., $J = z_{\mathrm{LAC}}(t_f)$, with $t_f$ = 30 h and 12 control actions ($N$ = 12). We furthermore consider box constraints for the inputs, namely $I_s = [0,1] \, \mathrm{W}/\mathrm{m}^{2}$ and $F_\textrm{in}=[0,1] \, \mathrm{L}/\mathrm{h}$. We add a further constraint to the optimization, $z_{\mathrm{GLC}}(t_f)= 0$, to ensure that all glucose, the feeding substrate, is fully consumed. Finally, the bioreactor volume should not surpass the maximum working volume capacity $v_{L_\mathrm{max}}$, thus we add the constraint $v_L \leq v_{L_\mathrm{max}}$.

\textit{Remark on the numerical solution of the optimization problem}: the solution to problems \eqref{eq:model_fed_batch}, \eqref{eq:op_con_fed_batch} and \eqref{eq:optimal_mpc} are optimal input \emph{functions}. This renders these problems infinite-dimensional, hence generally impractical to solve. One way to obtain a solution is via a finite-dimensional approach \citep{findeisen_introduction_2002,rawlings_model_2020}. In our case, this is achieved by assuming piece-wise constant inputs and by discretizing the ordinary differential equations using orthogonal collocation based on Lagrange interpolation polynomials as motivated by \cite{waldherr_dynamic_2015}. The bilevel optimizations in \eqref{eq:op_con_fed_batch}, \eqref{eq:optimal_mpc} and \eqref{eq:optimal_mhe} were transformed into mathematical programs with complementarity constraints (single-level optimizations) by applying the Karush-Kuhn-Tucker conditions to the inner optimization problems \citep{dempe_solution_2019}. The resulting optimizations were solved in Python using CasADi \citep{andersson_casadi_2019} and IPOPT \citep{wachter_implementation_2006}. 

\begin{table} [htb]
\begin{center}
\begin{threeparttable}
\caption{Relevant parameters and initial conditions of the nominal model.}\label{tb:optogenetic_parameters}
\begin{tabular}{cccc}
\toprule
Item & Value & Unit & Ref./Note \\\midrule
$\delta$ & 2.490 & 1 & \cite{olson_characterizing_2014} \\
$K$ & 0.138 & $\mathrm{W}  / \mathrm{m}^{2}$ & \cite{olson_characterizing_2014} \\ 
$\alpha$ & $2 \cdot 10^{-6}$ & $\mathrm{mmol} / \mathrm{g} / \mathrm{h}$ & Note 1 \\
$\beta$, S1 & $1 \cdot 10^{-4}$ & $\mathrm{mmol} / \mathrm{g} / \mathrm{h}$ & Note 1 \\
$\beta$, S2 & $2.5 \cdot 10^{-5}$ & $\mathrm{mmol} / \mathrm{g} / \mathrm{h}$ & Note 1 \\
$\beta$, S3 & $1 \cdot 10^{-5}$ & $\mathrm{mmol} / \mathrm{g} / \mathrm{h}$ & Note 1 \\
$d_{\mathrm{ATPase}}$ & $6.3 \cdot 10^{-2}$ & $1/\mathrm{h}$ & Note 2 \\
$l$ & $0.022$ & $\mathrm{m}$ & Note 3 \\
$v_{L_\mathrm{max}}$ & $45$ & $\mathrm{L}$ & Note 3 \\
$a_{\lambda}$ & $1 \cdot 10^{-2}$ & $\mathrm{m}^2 / \mathrm{g}$ & Note 4 \\
$z_{\mathrm{GLC,in}}$ & 2220 & $\mathrm{mM}$ & ----- \\
$x_{\mathrm{GLC}}(0)$ & 139 & $\mathrm{mM}$ & ----- \\
$x_{\mathrm{LAC}}(0)$ & 0 & $\mathrm{mM}$ & -----\\ 
$x_{\mathrm{CO}_2}(0)$ & 0 & $\mathrm{mM}$ & -----\\ 
$x_{\mathrm{FOR}}(0)$ & 0 & $\mathrm{mM}$ & -----\\ 
$x_{\mathrm{SUCC}}(0)$ & 0 & $\mathrm{mM}$ & -----\\ 
$B(0)$ & 0.59 & $\mathrm{g} / \mathrm{L}$ & -----\\
$\bm{p}(0)$ & Note 5 & $\mathrm{mM}$ & -----\\ 
$v_L(0)$ & 30 & $\mathrm{L}$  & -----
 \\\bottomrule
\end{tabular}
\small{\textbf{Note 1}. Assumed biologically sound values inferred from feasible deFBA simulations \citep{espinel_ATP_w_2022} for different induction strength scenarios (S\textsubscript{i}). \textbf{Note 2}. Estimated as $d_{\mathrm{ATPase}}=\frac{\ln (2)}{t_{0.5}}$, where $t_{0.5}$ is the ATPase protein half-life time \citep{benito_half-life_1991}. \textbf{Note 3}. Based on a pilot-scale flat-panel photobioreactor design \citep{koller_studies_2018}.
\textbf{Note 4}. Assumed biologically sound order of magnitude. Estimated as ca. $1/30$ of typical parameter values for microalgae \citep{blanken_predicting_2016}. \textbf{Note 5}. Estimated from $B(0)$ using resource balance analysis \citep{jabarivelisdeh_adaptive_2020}.}
\end{threeparttable}
\end{center}
\end{table}

\subsection{Open-loop optimal optogenetic control}
\label{sec:open_loop_results}
\begin{figure*}
\centering
\includegraphics[scale=0.70]{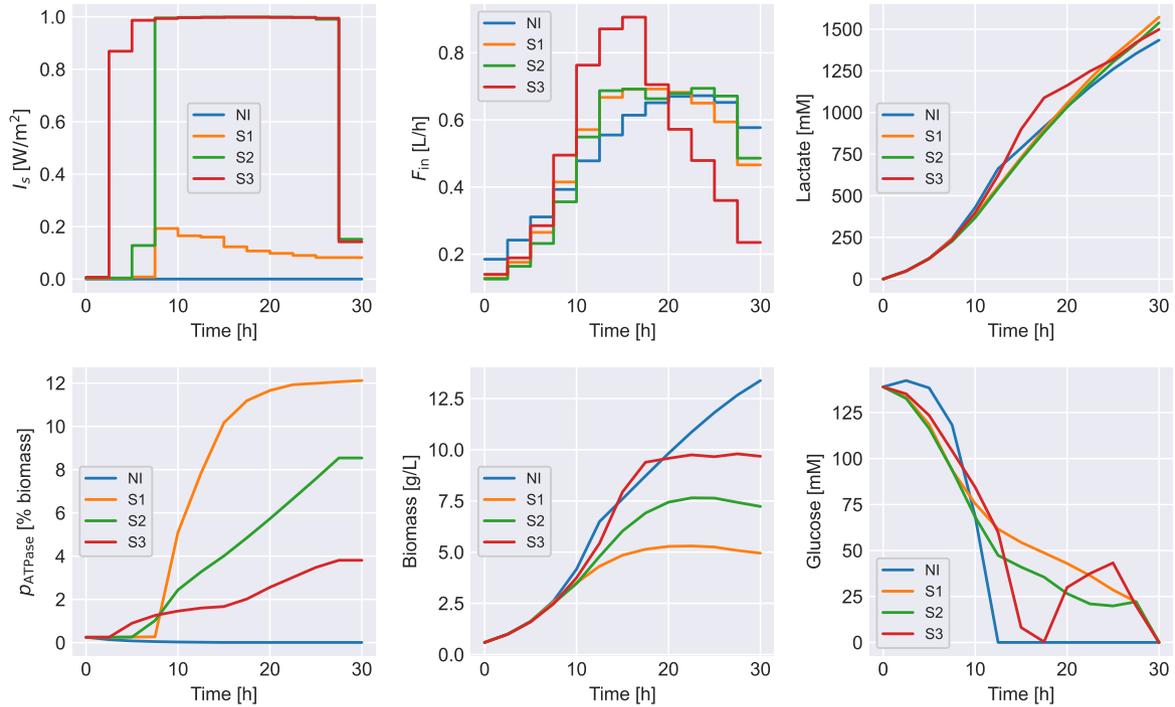}
\caption{Open-loop fed-batch simulations \textit{without} model-plant mismatch. Scenarios: NI--no induction, S1--high-strength, S2--medium-strength and  S3--low-strength inducible CcaS/CcaR systems.} \label{fig:fedbatch_results}
\end{figure*}

Fig. \ref{fig:fedbatch_results} shows the open-loop optimization results for the fed-batch fermentation considering \textit{no model-plant mismatch}. We depict four scenarios:
\begin{enumerate}
\item S1: high-strength inducible CcaS/CcaR system--high $\beta$ value. 
\item S2: medium-strength inducible CcaS/CcaR system--medium $\beta$ value.
\item S3: low-strength inducible CcaS/CcaR system--low $\beta$ value.
\item NI: no inducible ATPase enzyme--the CcaS/CcaR system is absent.
\end{enumerate}

\begin{figure*}
    \centering
    \begin{tikzpicture}
    \begin{axis}[
    name=fedbatch_bc,
    height=6.3cm,
    width=.26\linewidth,
    enlargelimits=false,
    axis on top,
    colormap/bluered,
    point meta=explicit,
    point meta min=0,
    point meta max=12.12,
    xtick={0,2,4,6,8,10,12,14,16,18,20,22,24,26,28,30},
    xticklabels={0,,,6,,,12,,,18,,,24,,,30},
    xlabel={\fontsize{0.3cm}{5.5cm}{\sffamily{Time [h]}}},
    ytick={100,150,200,250,300,350,400,450,500,550,600,650,700,750,800,850},
    yticklabels={$\mathrm{ptsGHI\_crr}$,$\mathrm{pgi}$,$\mathrm{pfkA\_fbaA}$,$\mathrm{tpiA}$,$\mathrm{gapA\_pgk}$,$\mathrm{gpmA\_eno}$,$\mathrm{pykF}$,$\mathrm{ldhA}$,$\mathrm{ATPase}$,$\mathrm{ppc}$,$\mathrm{pflB}$,$\mathrm{gltA\_acnB\_icd}$,$\mathrm{gdhA\_glnA}$,$\mathrm{mdh}$,$\mathrm{fumB}$,$\mathrm{frdABCD}$},
    yticklabel style = {font=\fontsize{0.3cm}{5.5cm},xshift=0.5ex},
    xticklabel style = {font=\normalsize}
    ]
        \addplot [matrix plot*] table [meta=concentration] {figures/enzymes_percentagema_BC.dat};
    \end{axis}
    
    \begin{axis}[
    name=fedbatch_s1,
    at=(fedbatch_bc.right of south east),
    anchor=left of south west,
        height=6.3cm,
    width=.26\linewidth,
    enlargelimits=false,
    axis on top,
    colormap/bluered,
    point meta=explicit,
    point meta min=0,
    point meta max=12.12,
    xtick={0,2,4,6,8,10,12,14,16,18,20,22,24,26,28,30},
    xticklabels={0,,,6,,,12,,,18,,,24,,,30},
    xlabel={\fontsize{0.3cm}{5.5cm}{\sffamily{Time [h]}}},
    ytick={100,150,200,250,300,350,400,450,500,550,600,650,700,750,800,850},
    yticklabels={,,,,,,,,,,,,,,,}
    ]
        \addplot [matrix plot*] table [meta=concentration] {figures/enzymes_percentagema_S1.dat};
    \end{axis}
    
    \begin{axis}[
    name=fedbatch_s2,
    at=(fedbatch_s1.right of south east),
    anchor=left of south west,
    height=6.3cm,
    width=.26\linewidth,
    enlargelimits=false,
    axis on top,
    colormap/bluered,
    point meta=explicit,
    point meta min=0,
    point meta max=12.12,
    xtick={0,2,4,6,8,10,12,14,16,18,20,22,24,26,28,30},
    xticklabels={0,,,6,,,12,,,18,,,24,,,30},
    xlabel={\fontsize{0.3cm}{5.5cm}{\sffamily{Time [h]}}},
    ytick={100,150,200,250,300,350,400,450,500,550,600,650,700,750,800,850},
    yticklabels={,,,,,,,,,,,,,,,}
    ]
        \addplot [matrix plot*] table [meta=concentration] {figures/enzymes_percentagema_S2.dat};
    \end{axis}

    \begin{axis}[
    name=fedbatch_s3,
    at=(fedbatch_s2.right of south east),
    anchor=left of south west,
    height=6.3cm,
    width=.26\linewidth,
    enlargelimits=false,
    axis on top,
    colormap/bluered,
    colorbar,
    colorbar style={ylabel=\fontsize{0.3cm}{5.5cm}{$\%$ \sffamily{biomass}}},
    point meta=explicit,
    point meta min=0,
    point meta max=12.12,
    xtick={0,2,4,6,8,10,12,14,16,18,20,22,24,26,28,30},
    xticklabels={0,,,6,,,12,,,18,,,24,,,30},
    xlabel={\fontsize{0.3cm}{5.5cm}{\sffamily{Time [h]}}},
    ytick={100,150,200,250,300,350,400,450,500,550,600,650,700,750,800,850},
    yticklabels={,,,,,,,,,,,,,,,}
    ]
        \addplot [matrix plot*] table [meta=concentration] {figures/enzymes_percentagema_S3.dat};
    \end{axis}
    
\end{tikzpicture}
    \caption{Enzyme expression heat map relative to the biomass dry weight for the fed-batch fermentations in Fig. \ref{fig:fedbatch_results}. From left to right, NI--no induction, S1--high-strength, S2--medium-strength and  S3--low-strength inducible CcaS/CcaR systems.}
    \label{fig:density_fedbatch}
\end{figure*}
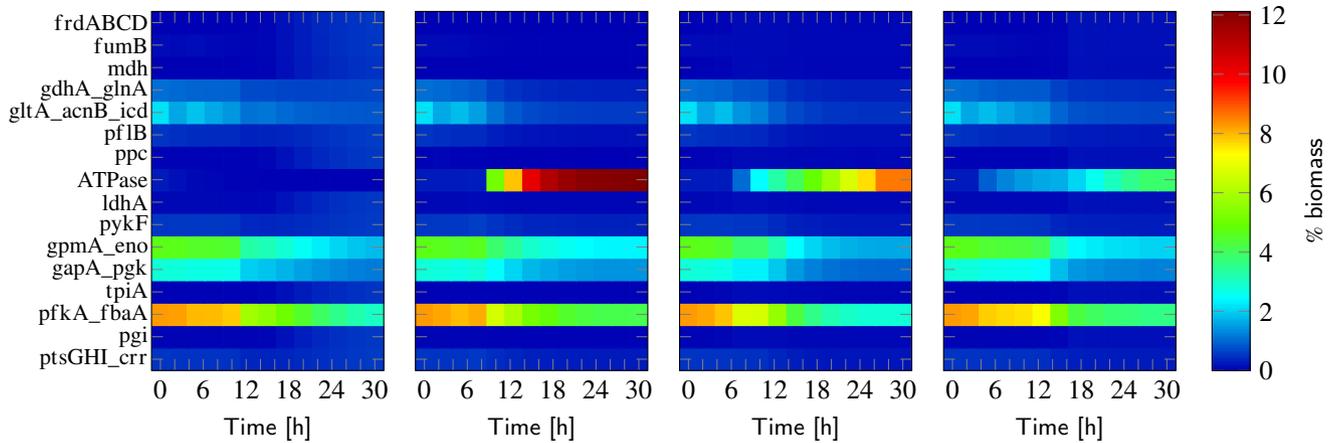

The NI case rendered a final lactate concentration of $1434.3\,\mathrm{mM}$, whereas S1 achieved $1572.3\,\mathrm{mM}$ ($\uparrow 10\,\%$), S2 $1538.5\,\mathrm{mM}$ ($\uparrow 7\,\%$) and S3 $1498.4\,\mathrm{mM}$ ($\uparrow 4\,\%$). Note that, by the end of all fermentations, the maximum allowed bioreactor volume was reached and all the glucose was fully depleted. This implies that overall the same net amount of glucose was fed and consumed. Consequently, in the previous scenarios, the relative gains in product titer also correspond, proportionally, to increments in product yield and volumetric productivity.

Furthermore, in S1 the maximum ATPase enzyme concentration was $12.1\,\%$ of the cell dry weight, in contrast to $8.5\,\%$ and $3.8\,\%$ in S2 and S3, respectively. As foreseeable, the higher the induction capacity strength of the CcaS/CcaR system, the higher the net ATPase enzyme expression, and therefore the higher the net increase in product yield. In previous works dealing with dynamic ATP turnover in one-stage batch fermentations, the increase in product yield was correlated with a loss in volumetric productivity \citep{espinel_ATP_w_2022,espinel_opt_2022}. Here we show that with a fed-batch system it is possible to increase both the product yield via the ATP turnover mechanism \textit{and} the volumetric productivity via the introduction of a feeding rate.

Compared to NI, the scenarios S1, S2 and S3 resulted in 63, 46 and $28\,\%$ lower final biomass concentrations. This can be explained by the combined effect of the lower biomass yields due to the ATP turnover and the potential resource burden related to the cost of producing the ATPase enzyme. Note that there is also a dilution effect from the feeding of the substrate. Overall, the increased ATP turnover rates managed to enhance the final lactate titer despite the lower biomass growth rates. 

In all induction scenarios, there was at first a gradual increase in the feeding rate, followed by a continuous decrease after around the mid-term of the fermentation. This allowed for making up sufficient biomass while keeping low induction levels of the ATPase enzyme. Then, to avoid excessive dilution of the biomass, the feeding rate decreased at increasing ATPase expression levels. 

A benefit of our model-based optimization is that it takes into account resource allocation constraints. The resource allocation phenomena associated with the expression of the ATPase enzyme is presented in Fig. \ref{fig:density_fedbatch}, where we show the dynamic enzyme composition profiles throughout the open-loop fermentations. Note that the induction of the ATPase enzyme led to a re-accommodation of the unregulated enzymes. For instance, let us compare the profiles of enzymes $\mathrm{frdABCD}$, $\mathrm{fumB}$, $\mathrm{mdh}$, $\mathrm{ppc}$, $\mathrm{ldhA}$, $\mathrm{tpiA}$ and $\mathrm{pgi}$. While they seem to slightly accumulate in the NI fermentation, they are kept at lower concentrations in the ATPase induction cases. The effect is clearer in scenario S1 because there the ATPase was expressed at higher levels.

\begin{figure*} [htb]
\centering
\includegraphics[scale=0.70]{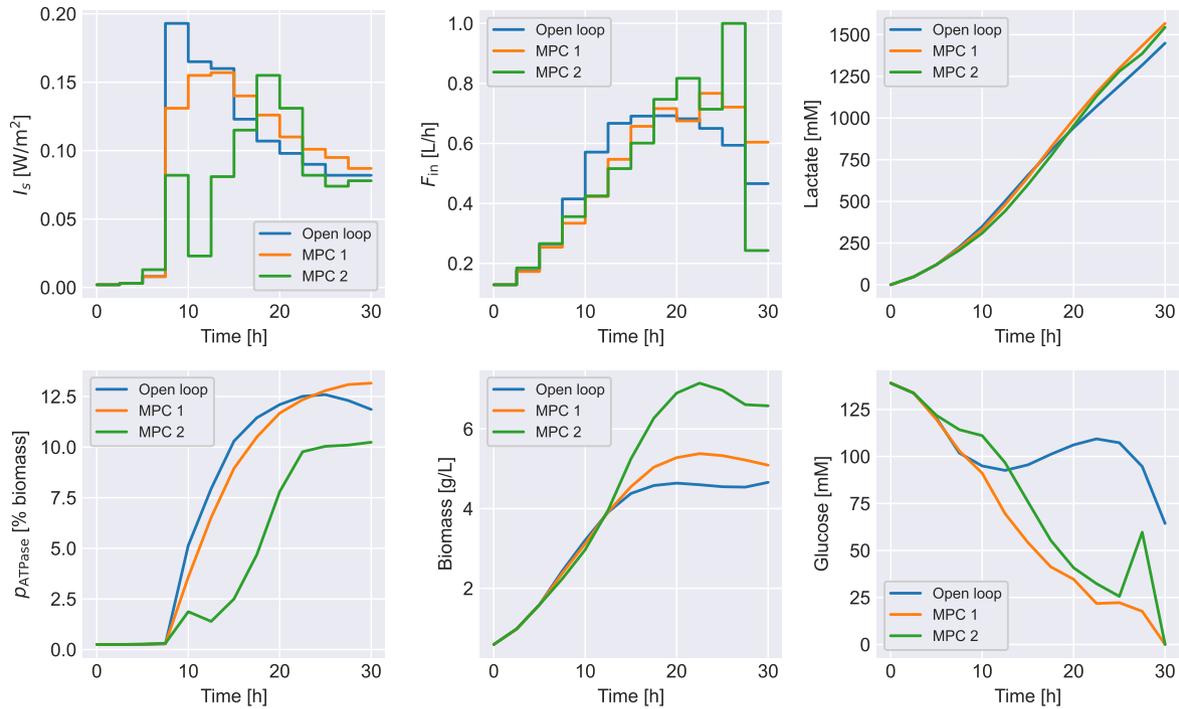}
\caption{Closed-loop fed-batch simulations with model uncertainty for the high-strength inducible CcaS/CcaR system. MPC 1: with full state measurement and no measurement noise. MPC 2: with measurements of $p_{\mathrm{ATPase}}$, $B$ and $\bm{z}$ in the presence of measurement noise; full information estimator used for estimating $\bm{p}$. The open-loop case (without online corrective actions) is also shown.} 
\label{fig:fedbatch_results_mpc}
\end{figure*}

\begin{figure*}[htb]
\subfloat{\includegraphics[width=0.25\textwidth, keepaspectratio]{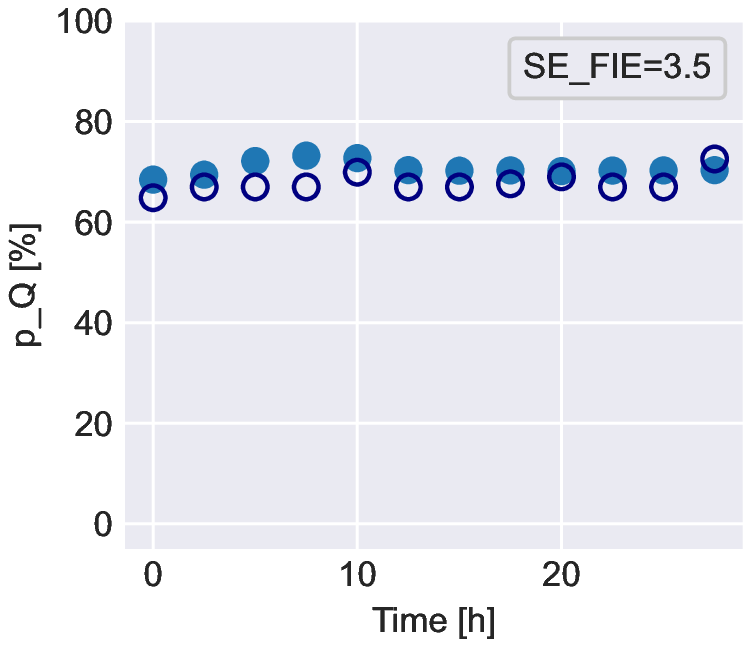}}
\subfloat{\includegraphics[width=0.25\textwidth, keepaspectratio]{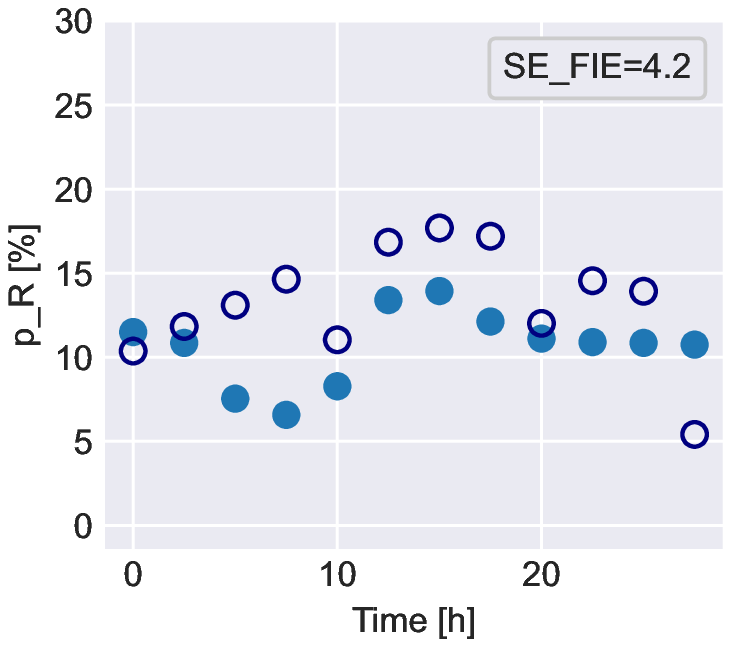}}
\subfloat{\includegraphics[width=0.25\textwidth, keepaspectratio]{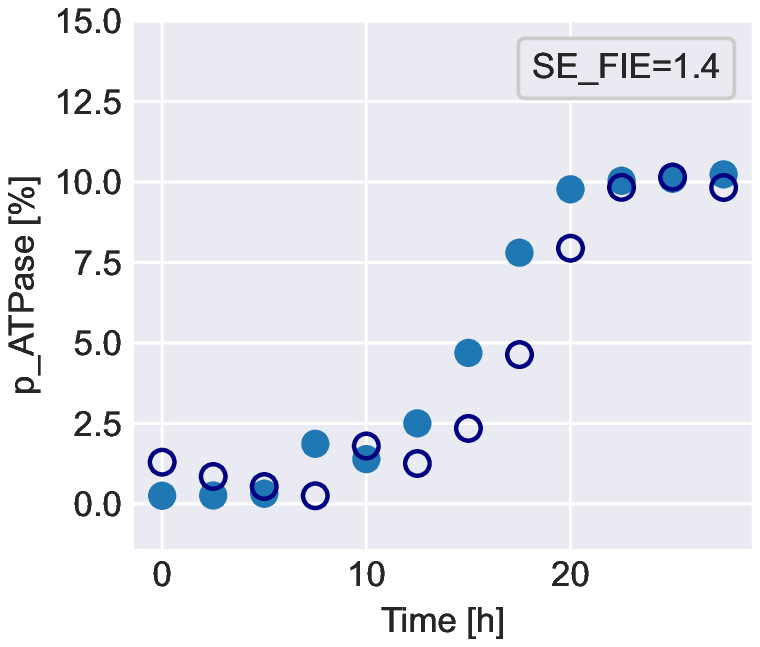}}
\subfloat{\includegraphics[width=0.25\textwidth, keepaspectratio]{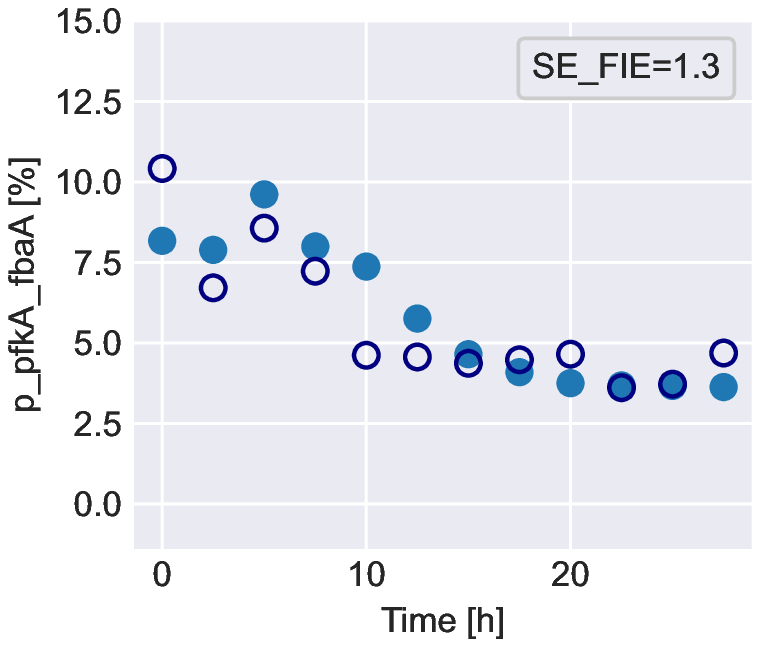}}
\\[-0.20cm]
\subfloat{\includegraphics[width=0.25\textwidth, keepaspectratio]{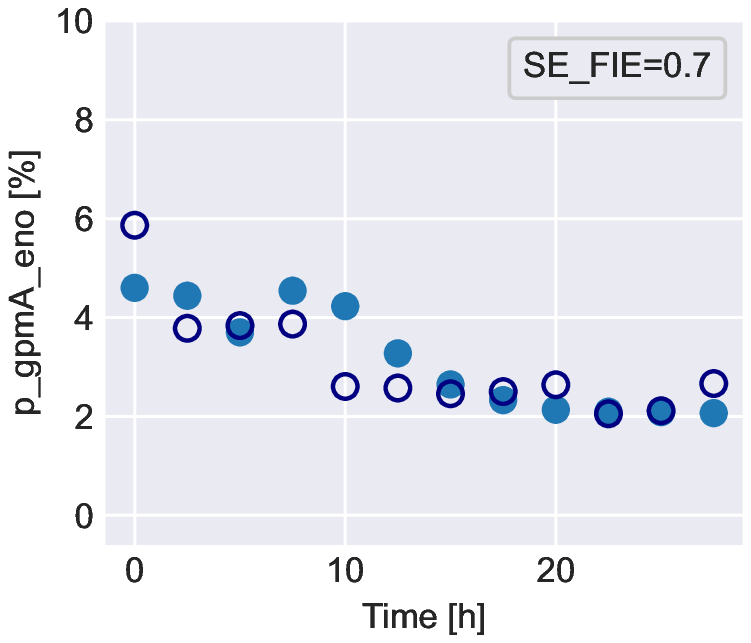}}
\subfloat{\includegraphics[width=0.25\textwidth, keepaspectratio]{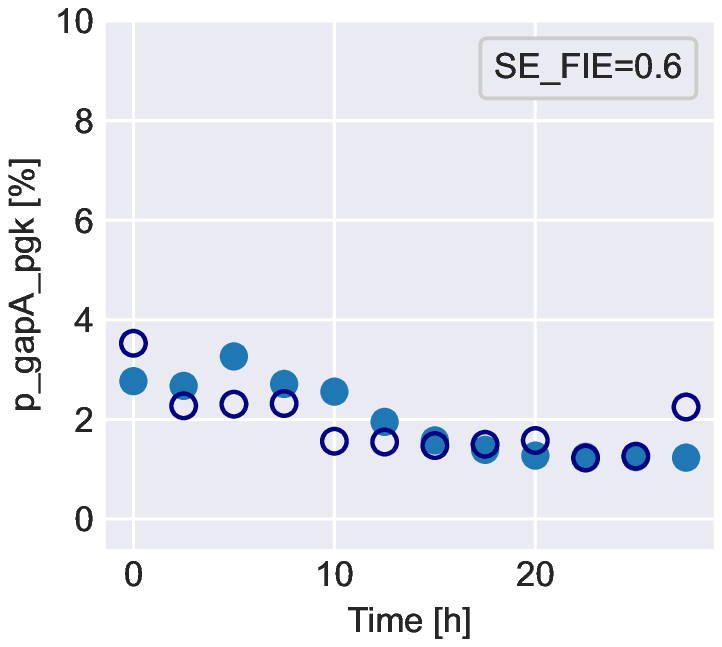}}
\subfloat{\includegraphics[width=0.25\textwidth, keepaspectratio]{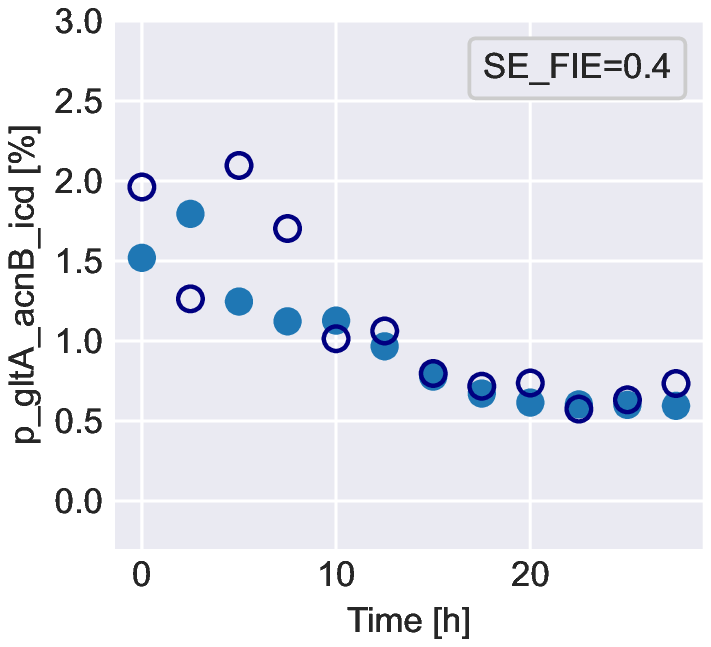}}
\subfloat{\includegraphics[width=0.25\textwidth, keepaspectratio]{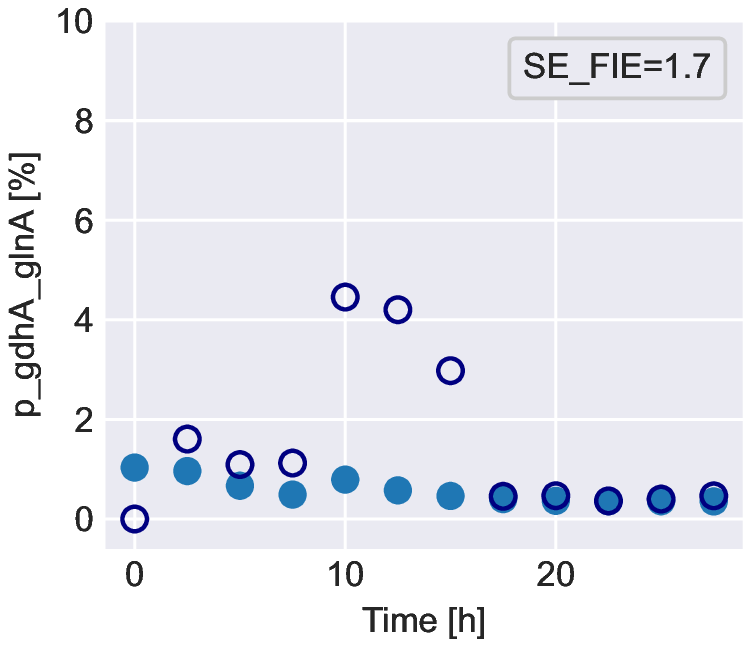}}
\caption[]{Online estimation of the cell components in percentage of cell dry weight. Only the eight most abundant species are shown. Filled circle: exact value. Empty circle: estimated state. The standard error of the estimate (SE\_FIE) is presented.}
\label{fig:estimation}
\end{figure*}

\subsection{Counteracting model uncertainties and disturbances -- optogenetic closed-loop control}
\label{sec:MPC_results}
Open-loop control does not account for model uncertainties and unforeseen disturbances and process changes. Thus, we now evaluate the performance of shrinking horizon MPC for addressing system uncertainty. We limit our analysis to the high-strength inducible CcaS/CcaR system which provided the best results. We introduced model-plant mismatch by scaling the catalytic constants of the enzymes pfkA\_fbaA, gpmA\_eno, gapA\_pgk, gltA\_acnB\_icd and gdhA\_glnA by a factor of $0.98$, which slightly decreases the fermentation rates. We also scaled down $\delta$ and $d_{\mathrm{ATPase}}$ by 0.97 and 0.98, respectively; the latter decreases the steepness of the Hill function and the former makes the ATPase enzyme slower-degrading. The \textit{modified} model was used for the plant simulations while the nominal model was given to the controller. Two MPC cases are considered:
\begin{enumerate}
\item MPC 1: all the states can be measured online without measurement noise. 
\item MPC 2: the concentrations of the ATPase enzyme\footnote{The ATPase concentration could be measured, e.g., using a fluorescence-based biosensor \citep{kim_genetically_2021}.}, biomass dry weight and extracellular metabolites can be measured online. Gaussian white noise ($1\,\%$ standard deviation) is added to the measurements. The cell composition is estimated via full information estimation\footnote{For simplicity, in the full information estimation we only consider the second term of the objective function in \eqref{eq:obj_mhe}. We neglect the state noise and assume constant model parameters, hence they are not estimated. The matrix $\mathbf{R}$ is chosen as the identity matrix.}. The \textit{reconstructed} cell composition, along with the online measurements, are passed to the MPC. 
\end{enumerate}

The MPC simulations are shown in Fig. \ref{fig:fedbatch_results_mpc}. We also plot the open-loop scenario (\textit{with model-plant mismatch}) as a reference case. The open-loop controller resulted in a final lactate concentration of $1449.5\,\mathrm{mM}$ and $64.3\,\mathrm{mM}$ net unconsumed glucose. The applied light intensity brought the ATPase enzyme concentration up to $12.6\,\%$ of the cell dry weight, but then it decreased slightly to $11.9\,\%$.

The applied light intensity in MPC 1 allowed the ATPase enzyme fraction in the cell to eventually surpass the value achieved in the open-loop fermentation. The combined effect of the corrected light intensities and feeding rates rendered a final lactate titer of $1567.8\,\mathrm{mM}$. The latter represents an $8\,\%$ improvement with respect to the open-loop optimization. Also, with MPC 1 there was no unconsumed glucose by the end of the process. Note that MPC 1 scenario is a very optimistic result as the full state measurement is assumed and there is no measurement noise present. 

MPC 2 results are more realistic regarding practical implementation, i.e., with state estimation and measurement noise. The estimation of the cell composition for selected species at the different sampling times is presented in Fig. \ref{fig:estimation}. We also calculated the standard error (SE) of the estimates\footnote{$\mathrm{SE}_\mathrm{FIE}= \sqrt{\frac{\sum_{\substack{j={0}}}^{N-1} (p_{i,j}-{p^*}_{i,j})^2}{E_i}}, \, \forall i \in [1,\,n_{p_i}]$, $E_i$: total number of $p_i$ estimates.}. Overall, the full information estimator tracked well the concentration trends of the biomass components. However, it should be noted that, in general, the estimation improved as the process proceeded. That is, the estimations were less accurate during the first one-third of the process (cf. e.g. the estimation profiles of enzymes $\mathrm{pfkA\_fbaA}$, $\mathrm{gltA\_acnB\_icd}$ and $\mathrm{gdhA\_glnA}$). The progressive improvement of the estimation is explained by the growing estimation horizon and thus the increasing number of available measurements. This furthermore explains why at the beginning of the process the controller´s predictions were comparatively \textit{off} with respect to MPC  1 and the open-loop optimization.  MPC 2 reached an intracellular ATPase enzyme concentration of about $10.2\,\%$, leading to more biomass accumulation. The controller adjusted the feeding rates to avoid having unconsumed glucose by the end of the process. MPC 2 rendered a final lactate concentration of $1544.4\,\mathrm{mM}$, very close to the value achieved in the MPC 1 scenario. 

The MPC simulations demonstrate that using model-based feedback control, optionally coupled to state estimation methods, can improve the process performance of metabolic cybergenetic systems in the presence of system uncertainty.

\section{Conclusions and Outlook}
\label{sec:conclusions}
We propose to fuse cybergenetics with model-based optimization and predictive control for dynamic metabolic engineering applications. The proposed metabolic cybergenetic framework exploits the concept of online metabolic regulation by dynamically modulating the gene expression of metabolism-relevant intracellular proteins. To do so, we developed a dynamic constraint-based modeling framework that integrates the dynamics of metabolic reactions, resource allocation and external gene expression regulation. The model is combined with model-based optimization, predictive control and estimation methods to facilitate the implementation of metabolic cybergenetic systems.

The potential of this technology is highlighted considering the dynamic control of the cellular ATP turnover via optogenetic regulation of the ATPase gene expression. We show that optimal control of the light intensity and the substrate feeding rate can enhance the process performance in terms of product titer and volumetric productivity. Furthermore, we demonstrated that introducing feedback via model predictive control can help to counteract system uncertainty.

We believe that the outlined metabolic cybergenetic framework opens the door to new and more advanced biotechnological applications where manipulating metabolic fluxes throughout the process is required. This is actually in line with dynamic metabolic engineering approaches and goes beyond traditional cybergenetic schemes where the regulated proteins (e.g., fluorescence reporters) are not directly involved in metabolic pathways. Moreover, the model-based feature of the presented framework can contribute to shortening and reducing the cost of process development, and obtaining a more robust, consistent and flexible operation. 

We currently develop metabolic cybergenetic methods augmented with machine learning, also extending their scope to synthetic microbial communities. Finally, we aim to experimentally validate the proposed framework, considering the presented case study and other relevant bioprocesses.

\section{Acknowledgment}
This work was supported by the International Max Planck Research School for Advanced Methods in Process and Systems Engineering (IMPRS ProEng) and by the EU-EFRE funded project DIGIPOL.

\bibliography{bibliography}






\end{document}